\documentstyle[prl,aps,epsf]{revtex}

\begin{document}
\draft
\twocolumn[\hsize\textwidth\columnwidth\hsize\csname@twocolumnfalse\endcsname

\title{Evidence for a two component magnetic response in UPt$_3$}
\author{A. Yaouanc,$^1$ P. Dalmas de R\'eotier,$^1$ F.N. Gygax,$^2$
A. Schenck,$^2$ A. Amato,$^3$ C. Baines,$^3$ P.C.M. Gubbens,$^4$ 
C.T.~Kaiser,$^4$  A. de Visser,$^5$ 
R.J. Keizer,$^5$ A. Huxley$^1$ and A.A. Menovsky$^5$}

\address{$^1$Commissariat \`a l'Energie Atomique, D\'epartement de Recherche
Fondamentale sur la Mati\`ere Condens\'ee \\
F-38054 Grenoble Cedex 9, France}
\address{$^2$Institute for Particle Physics, Eidgn\"ossische Technische 
Hochschule Zurich, CH-5232 Villigen PSI, Switzerland}
\address{$^3$Laboratory for Muon Spectroscopy, Paul Scherrer Institute,
CH-5232 Villigen PSI, Switzerland} 
\address{$^4$Interfacultair Reactor Instituut, Delft University of Technology,
2629 JB Delft, The Netherlands}
\address{$^5$Van der Waals-Zeeman Instituut, Universiteit van Amsterdam, 
1018 XE Amsterdam, The Netherlands}
\date{\today} \maketitle 

\begin{abstract}
The magnetic response of the heavy fermion superconductor UPt$_3$ has been 
investigated on a microscopic scale by muon Knight shift studies.
Two distinct and isotropic Knight shifts have been found for the field in the 
basal plane. While the volume fractions associated with the two Knight shifts 
are approximately equal at low and high temperatures, they show a 
dramatic and opposite temperature dependence around $T_N$. Our results are 
independent on the precise muon localization site. We conclude that
UPt$_3$ is characterized by a two component magnetic response.

\end{abstract}
]
The hexagonal heavy fermion superconductor UPt$_3$ is attracting much interest
because it has been established as an unconventional superconductor
as seen by the existence of three  
distinct superconducting phases in the magnetic field-temperature plane
\cite{Sauls94,Heffner96}. In zero-field the two superconducting phase 
transitions occur at $\sim$ 0.475~K and $\sim$ 0.520~K. 
It is usually thought that this complex phase 
diagram arises from the lifting of the degeneracy of a 
multicomponent superconducting order parameter. 

The most popular candidate for such a symmetry-breaking field is the short 
range antiferromagnetic order characterized by a N\'eel temperature of 
$T_N$ $\simeq$ 6 K and an extremely small ordered magnetic moment 
(0.02 (1) $\mu_B$/U-atom in the limit $T \rightarrow 0$ K) oriented along 
the $a^*$ axis ($\equiv b$ axis). The magnetic order has only been 
observed by neutron \cite{Aeppli89} and magnetic x-ray \cite{Isaacs95}
diffractions. 

Nuclear magnetic resonance \cite{Tou} and zero-field muon spin relaxation
\cite{Dalmas95} measurements as well as macroscopic studies have failed to 
prove the existence of static antiferromagnetic order on high quality samples 
\cite{extra}. Here we present transverse high-field muon spin rotation 
($\mu$SR) data which present anomalies around $T_N$. Moreover, they 
show that UPt$_3$ is characterized by 
a two component magnetic response at least up to 115 K.

In the transverse $\mu$SR technique \cite{muon}, polarized muons are 
implanted into a sample where their spins ${\bf S}_{\mu}$ ($S_{\mu}$ = 1/2)
precess in the local magnetic field ${\bf B}_{\rm loc}$ until they decay.
The sample is polarized by a magnetic field ${\bf B}_{\rm ext}$ applied 
perpendicularly to  ${\bf S}_{\mu}(t=0)$. 
${\bf S}_{\mu}(t)$ is monitored through the decay positron.
By collecting several million positrons, one can readily obtain an accurate 
value for the field at the muon site(s).

We present results for three samples. Two samples have been grown in 
Grenoble. Each consists of crystals glued on a silver backing plate and put 
together to form a disk. They differ by the orientation of the crystal 
axes relative to the normal to the sample plane: either the 
$a^*$ or $c$ axis is parallel to that direction. Measurements have 
therefore been carried out either with ${\bf B}_{\rm ext}$ parallel to $a^*$
or $c$. The third sample has been prepared in 
Amsterdam. It is a cube of 5 $\times$ 5 $\times$ 5 mm$^3$ which has been 
glued to a thin silver rod. The measurements on this sample have 
been done only with ${\bf B}_{\rm ext}$ $\parallel a$.
The Grenoble samples have already been used for zero-field  \cite{Dalmas95} 
and transverse low-field \cite{Yaouanc98} $\mu$SR measurements. Their high 
quality is demonstrated by the splitting of the two zero-field
superconducting transitions as seen by specific heat \cite{Yaouanc98} 
and the low residual resistivities which are among the lowest ever reported 
($\rho_c(0)$ = 0.17 $\mu \Omega$~cm and $\rho_{a^*}(0)$ = 0.54 $\mu \Omega$~cm
\cite{Suderow98}). The Amsterdam sample is of a somewhat lesser
quality in terms of the residual resistivities which are roughly a
factor 3 higher than for the Grenoble sample. 
Nevertheless the double superconducting transition is clearly resolved in the 
specific heat.

The measurements have been performed at the low temperature facility
(LTF) and at the general purpose spectrometer (GPS) of the $\mu$SR facility 
located at the Paul Scherrer Institute. The 
LTF spectra have been obtained for temperatures between 0.05 K and 10 K and 
$B_{\rm ext}$ of 2.3 T (only for two measurements), 2 T and 1.5 T. 
The GPS data have been taken with 
$B_{\rm ext}$ = 0.6 T for 1.7 K $\leq$ $T \leq$ 200 K. A high 
statistic GPS measurement 
has been carried out at 50 K with $B_{\rm ext}$ = 0.45 T. The GPS 
measurements have been performed with an electrostatic kicker device 
on the beam line which ensures that only one muon at a time is present in 
the sample \cite{Abela99}. With such a device, the signal to noise ratio 
is strongly enhanced and the time window is extended to $\sim$ 18
$\mu$s. For both 
spectrometers ${\bf B}_{\rm ext}$ has been applied along the muon beam 
direction and a spin rotator has been used to flip the muon spin away from 
the muon momentum. 

We expect to observe a sum of oscillating 
signals, each corresponding to a given type of muon environment. 
An extra signal originating from muons stopped in the sample
surroundings, basically a silver backing plate, is also expected.

\begin{figure}
\epsfxsize=82 mm
\epsfbox{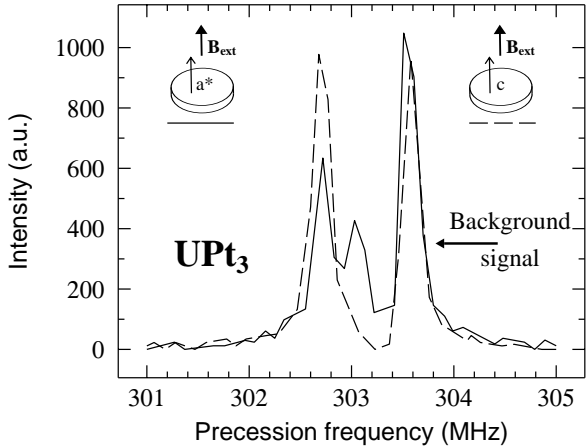}
\caption{Two Fourier transforms of spectra recorded at 2.6 K with 
$B_{\rm ext}$ = 2.3 T. ${\bf B}_{\rm ext}$ is either parallel to the $a^*$ 
or $c$ axis. The background signal was
intentionally enlarged in this measurement to evidence the difference
between the applied field and the field at the muon site. This was
achieved by fixing a 10 mm diameter Ag mask on the sample, 
resulting in a reduced effective sample size.
The line at $\sim$ 303.6 MHz originates from the Ag mask.}
\label{fig_fourier}
\end{figure}

\begin{figure}
\epsfxsize=82 mm
\epsfbox{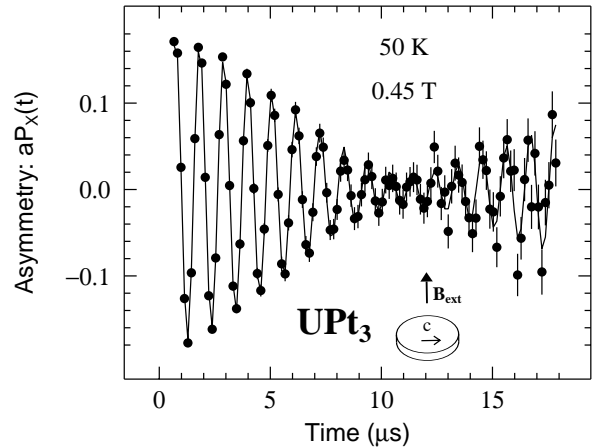}
\caption{A spectrum recorded at 50 K in a field of 0.45 T applied 
perpendicular to the $c$ axis of UPt$_3$ and presented in a 
frame rotating at a precession frequency of 60 MHz. This experiment 
was perfomed
in a setup designed to get the smallest possible background ($a_{\rm bg}
\simeq$ 0.017). 
The solid line is a fit 
to a sum of three oscillating components: two from the sample and one from the 
background. 
The reduced asymmetry observed on the plot around 10 $\mu$s, which  is too
small to be explained by a beating of the UPt$_3$ and background
signals, reflects the 
beating of the two signals originating from UPt$_3$. }    
\label{fig_spectra}
\end{figure}

In Fig.~\ref{fig_fourier} we present two Fourier transforms of spectra. 
Two lines from the sample are clearly detected for 
${\bf B}_{\rm ext}$ $\parallel a^*$. A symmetric single line is observed 
for ${\bf B}_{\rm ext}$ $\parallel c$. 
For the whole temperature range investigated two components are found for 
${\bf B}_{\rm ext}$ $\perp c$ and only 
a single component is detected for ${\bf B}_{\rm ext}$ $\parallel c$. 
In Fig.~\ref{fig_spectra} we present a time spectrum which clearly shows the 
existence of the two components far into the paramagnetic regime for 
${\bf B}_{\rm ext}$ $\perp c$.
    
We first discuss the spectra recorded for ${\bf B}_{\rm ext} \perp c $
which have been analyzed with the polarization function $P_X(t)$ written as 
the sum of three components:

\begin{eqnarray}
aP_X(t) & & = 
a_{\rm F} \cos \left( \omega_{\rm F} t \right) 
\exp \left (- {\Delta^2 t^2/  2} \right) 
\cr 
+ a_{\rm S} \cos & & \left (\omega_{\rm S} t \right) \exp(- \lambda t) 
+ a_{\text{bg}}\cos \left (\omega_{\text{bg}} t \right) 
\exp(- \lambda_{\text{bg}} t). 
\label{fit}
\end{eqnarray}

The first two components describe the $\mu$SR signal from the sample and the 
third accounts for the muons stopped in the background. The subscripts F and 
S refer to the first and second components, respectively. $a_{\alpha}$ is the 
initial asymmetry of component $\alpha$ oscillating at the pulsation frequency 
$\omega_{\alpha}$ = $2 \pi \nu_{\mu,\alpha}$ = 
$\gamma_\mu B_{{\rm loc}, \alpha}$ 
where $\nu_{\mu,\alpha}$ is the precession frequency of component $\alpha$
and $\gamma_\mu$ the muon gyromagnetic ratio 
($ \gamma_ \mu$ = 851.6 Mrad s$^{ {\rm -1}}$T$^{-1} $). 
$a_\alpha$ is proportionnal to the fraction of muons experiencing field
$B_{{\rm loc}, \alpha}$.
The envelop of the first component is 
best fitted by a Gaussian function, while the envelop of the second component
is better described by an exponential damping. We stress that the measured
temperature dependences of the two initial asymmetries and frequencies are not 
influenced by the choice of the envelop functions.
 
$\Delta$ is approximately independent of the temperature and amounts to
$\simeq$ 0.55 MHz at high field. It roughly scales with $B_{\rm ext}$.
$\lambda$ is independent of $B_{\rm ext}$ and is equal to 
$\lambda$ $\simeq$ 0.14 MHz at the lowest temperature. It decreases when the 
temperature is increased and becomes so small above 4 K that it can
be fixed to zero. The values
of the damping rates may reflect only partially the intrinsic properties of 
UPt$_3$ because of the field inhomogeneity due to the demagnetization field.  
However, $\Delta$ and $\lambda$ are remarkably small, indicating that the 
magnetic inhomogeneity detected for the two components is small.

\begin{figure}
\epsfxsize=82 mm
\epsfbox{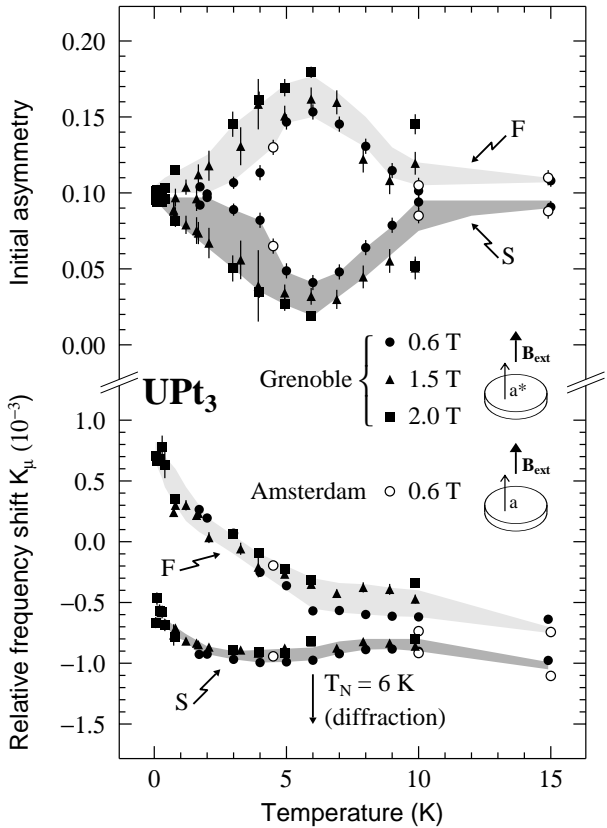}
\caption{Temperature dependence of the initial asymmetries and Knight 
shifts $K_{\mu}$
with ${\bf B}_{\rm ext}$ perpendicular to the $c$ axis. The data are for 
0.5 K $\leq$ $T$ $<$ 15 K, three field intensities and two samples denoted as 
Grenoble and Amsterdam. $K_{\mu}$ is corrected for Lorentz and 
demagnetization fields. The actual value of $K_\mu$ is subject to an 
uncertainty due to the demagnetization correction. Nevertheless the shape  
of $K_\mu(T)$ is independent on this correction.
The F and S letters denotes the two components.}
\label{parametre_astar}
\end{figure}

In Fig.~\ref{parametre_astar} we display the temperature dependence of the two
initial asymmetries and the associated relative frequency shifts, $K_{\mu}$. 
These plots 
concern the spectra taken with ${\bf B}_{\rm ext} \perp c$ and for 
$T \leq$ 14.7 K. $K_{\mu}$, which is the local magnetic susceptibility at the 
muon site, is usually called the Knight shift. It is deduced from the 
measured relative frequency shift, $K_{\rm exp}$, after correcting for the 
Lorentz and demagnetization fields. $K_{\rm exp}$ is 
defined by $K_{\rm exp}$ $= {\bf B}_{\rm ext} \cdot ({\bf B}_{\rm loc} 
- {\bf B}_{\rm ext})/ B_{\rm ext}^2$. We have determined $B_{\rm ext}$ with 
a gaussmeter or through the pulsation frequency 
of the background: $B_{\rm ext}$ = $\omega_{\text{bg}}/ \gamma_\mu$.
Since the Knight shift of the background is very small
($K_{\mu}$ for silver is 
$\simeq$ 94 ppm \cite{Schenck82}), this is a very good approximation. 
Although the Lorentz and demagnetization correction modifies substantially 
the absolute value of the Knight shift, qualitatively it does not influence
its temperature dependence. The conclusions we shall draw from our 
data are independent of the uncertainty related to the correction. 

The results of Fig.~\ref{parametre_astar} show that the  
Grenoble and Amsterdam samples yield consistent results. Since, as 
indicated in the figure, the measurements have been done either with
${\bf B}_{\rm ext}$ $\parallel a^*$ or ${\bf B}_{\rm ext}$ $\parallel a$, 
we conclude that the $\mu$SR response is isotropic 
in the basal plane. The data display also two remarkable features.

First, the frequency splitting between the two lines is relatively large 
at low $T$, decreases rapidly as $T$ increases up to $\sim$ 6 K and exhibits 
a shallow minimum around 10 K. It increases again for higher temperatures
(not shown). This explains the possibility of observing the beating between 
the two oscillating components at 50 K as shown in Fig.~\ref{fig_spectra}. 
Only the F shift has a 
strong temperature dependence below 6 K while the S shift is practically 
temperature independent down to 0.4 K, below which its absolute value slightly 
decreases. Since 6 K is the $T_N$ value as determined by neutron 
diffraction on our samples, the temperature dependence of the F shift provides
a signature of the N\'eel temperature. 

The second feature is probably the most striking: we observe two muon 
precession frequencies with approximately equal initial asymmetries in the 
whole temperature 
range (0.05 K $\leq$ $T \leq$ 200 K, the region $T >$ 15 K is not shown in 
Fig.~\ref{parametre_astar}) except near $T_N$ ($T_N$ $\pm$ 4 K) 
where $a_{\rm F}$ increases at the expense of $a_{\rm S}$. 

In this temperature range a
trend for a larger difference between these initial asymmetries seems to be
present at high field. However this trend 
might not be meaningful since the signal to noise ratio for 
the 2.0 T and 1.5 T spectra is not as good as for the 0.6 T spectra. An 
eventual field effect on $a_{\rm F}$ and $a_{\rm S}$ could only be 
confirmed by 
measurements with the electrostatic kicker device at all fields. Since
high-field neutron diffraction \cite{Lussier96,Dijk98} did not detect
any sizeable change in the
relative population of the three equivalent antiferromnagnetic domains 
we do not expect a field effect on the initial asymmetries.

The spectra recorded with ${\bf B}_{\rm ext} \parallel c $ have been analysed 
with a formula similar to Eq.~\ref{fit} with $a_{\rm S}$ = 0. The precession
frequency varies smoothly in temperature. 
The Gaussian damping rate scales again with $B_{\rm ext}$ ($\sim$ 0.42
MHz at 1.5 T) and is essentially temperature independent up to
$\sim$ 30 K above which temperature it drops smoothly to very
small values.

\begin{figure}
\epsfxsize=82 mm
\epsfbox{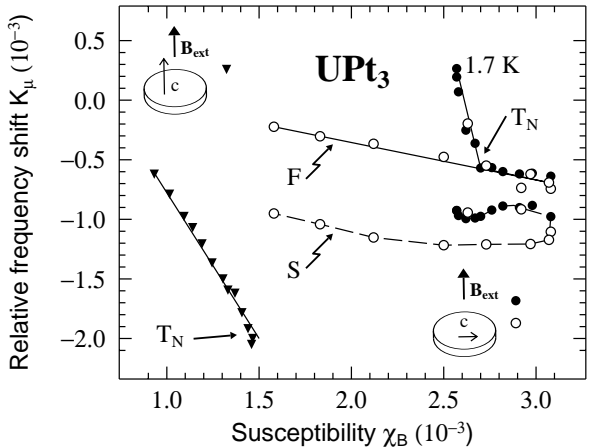}
\caption{Clogston-Jaccarino plots obtained for $B_{\rm ext}$ = 0.6 T. 
For ${\bf B}_{\rm ext} \perp c$ we use the same symbol convention as in 
Fig. \ref{parametre_astar}. The filled triangles correspond to ${\bf
B}_{\rm ext} \parallel c$. The temperature is an implicit parameter: 
1.7 K $\leq$ $T \leq$ 115 K. $K_\mu$ is extracted from the reported 
$\mu$SR measurements and the bulk susceptibility $\chi_B$ from measurements 
on our samples. The lines are guides to the eyes. Note the drastic change of 
slope at $T_N$ for the F set of data. }
\label{jaca}
\end{figure}

In Fig.~\ref{jaca} we present the $K_{\mu}$ data recorded for 
$B_{\rm ext}$ = 0.6 T with 1.7 K $\leq T \leq$ 115 K as a function of the 
bulk susceptibility $\chi_B$. This is a so called Clogston-Jaccarino plot, the 
temperature is an implicit parameter. The bulk susceptibilities for the 
different orientations have been measured on the Grenoble samples and are 
similar to those of Ref.~\cite{Frings83}. Classically, we should find 
$K_{\mu}$ scaling with the susceptibility. This is approximately observed for 
${\bf B}_{\rm ext} \parallel c$ but not for ${\bf B}_{\rm ext} \perp c$. 
In addition, as already pointed out when discussing $K_{\mu}(T)$, the 
Clogston-Jaccarino plots clearly show that while the F Knight shift provides a 
signature of $T_N$, such a signature is absent for the S Knight shift. 
The data of Fig.~\ref{jaca} suggest that $K_\mu$ passes smoothly through $T_N$
for ${\bf B}_{\rm ext} \parallel c$, although the almost constant value of 
$\chi_B(T)$ at low temperature does not allow for a definite statement.

We now discuss the muon diffusion properties and localization site in UPt$_3$. 
The shape of the zero-field depolarization and the constant value of the 
related damping rate show that the muons are static and occupy the same site
in the muon time scale, at least below 30 K \cite{Dalmas95}. 
Our transverse field measurements suggest that in fact the muon is diffusing 
only above 115 K because the frequency splitting collapses above that 
temperature (not shown). Since we focus on the properties of UPt$_3$ itself, 
we only consider the data for which the muon is static. Thus the 
anomalous temperature dependence of the two initial asymmetries 
around $T_N$ for 
${\bf B}_{\rm ext} \perp c$ can not be due to muon diffusion. The 
eventual existence of two distinct muon sites can not explain our data
since their relative occupancy should not change for a static muon. 
Interestingly, the analysis for 
U(Pt$_{0.95}$Pd$_{0.05}$)$_3$ of the angular dependence of $K_{\mu}$ shows
that the muon occupies only one site, the 2a site in Wyckoff notation 
(P6$_3$/mmc space group) in this related compound
\cite{Schenckfuture}.

Our results are understood if we suppose that the muon 
occupies only one magnetic site and the sample is intrinsically 
inhomogeneous: it consists of two regions with slightly different magnetic 
responses and relative volumes which are temperature dependent. While near 
$T_N$ one region dominates, outside that temperature range the two regions 
occupy approximately equal volumes. An alternative explanation for our
data could be the existence of a complex magnetic structure leading to
the observed $\mu$SR response. However this would imply a more involved
magnetic structure than the one published 
\cite{Aeppli89,Isaacs95,Lussier96,Dijk98}. In addition it is difficult
to imagine that a magnetic structure can influence the muon response up 
to at least 20 $T_N$. Therefore we
disregard this latter explanation.

The facts that the magnetic phase transition is only detected by transverse 
high-field and not by zero-field $\mu$SR measurements 
\cite{Dalmas95,Visser97b} 
are not inconsistent. It is not unexpected to observe below $T_N$ a new 
source of quasi-static magnetic polarization induced by the applied field 
which leads to an extra Knight shift. 

Bulk magnetic susceptibility does not detect the phase transition since the 
relative sensitivity in these conditions is $\simeq$ 10$^{-3}$. As shown in 
Fig.~\ref{parametre_astar}, this is not enough.

The results obtained by the $\mu$SR and magnetic diffraction techniques are 
not contradictory. The diffraction results simply mean that the
difference in the scattering properties of the two regions may be too subtle
to be distinguished.

We now consider the possible origin for the additional Knight shift observed 
below $T_N$ for the F component. A change of the magnitude of the 
moments is excluded since 
high-field neutron diffraction measurements do not detect any sizeable 
influence of a field up to 12 T \cite{Lussier96,Dijk98}. Two  
mechanisms producing an additional shift can be imagined. 
The first mechanism involves the dipolar field produced at the muon site by 
the ordered uranium moments. A rotation of these moments induced by the 
applied field leads to an additionnal field at the muon site. A small 
rotation is not excluded by neutron diffraction since this technique gives an 
upper bound rotation angle as large as 26$^\circ$ \cite{Lussier96}.  
But it is surprising, for a magnet with moments oriented along the $a^*$ 
direction to observe the same $K_{\mu}$ for
${\bf B}_{\rm ext} \parallel a$ and ${\bf B}_{\rm ext} \parallel a^*$
(see Fig.~\ref{parametre_astar}). The second possible origin focuses on the 
itinerant character of the magnetism of UPt$_3$. 
In this picture the additional shift is a measure of the enhancement of the 
magnetic susceptibility of the conduction electrons below $T_N$. UPt$_3$ 
being a planar magnet with a negligible planar anisotropy, the enhancement 
should be isotropic in the plane perpendicular to $c$ and no enhancement 
should be observed for ${\bf B}_{\rm ext} \parallel c$. This is consistent 
with our data.

Our most surprising result is the existence of the two components when 
${\bf B}_{\rm ext} \perp c$. Since the associated damping rates are  
small, we infer that the magnetic disorder is small. The near equality in 
most of the temperature range of the two initial asymmetries suggests that 
the two regions originate from a periodic modulation. The behaviour
of the initial asymmetries near $T_N$ implies that 
the proposed modulation is strongly coupled to the magnetic order parameter. 
The structural modulation observed by electron microscopy and  
diffraction some years ago \cite{Midgley93} might be related to the 
regions discussed here. However it has never been seen 
thereafter including in our samples.

In summary we have discovered by transverse high-field $\mu$SR measurements
the existence of a two component magnetic response. While the volume
fraction associated with these components is equal below 
$\sim$ 2 K and above $\sim$ 10 K, it is strongly temperature dependent 
around $T_N$. 
We also observe a signature of the 
magnetic transition for one of the two components. 
Our results are naturally explained if UPt$_3$ is
intrinsically inhomogeneous at least in a applied field.

\vspace{-.5  cm} 


\begin{references}
\vspace{-1.2 cm} 


\bibitem{Sauls94} J.A. Sauls, Adv. in Phys. {\bf 43}, 113 (1994).
\bibitem{Heffner96} R.H. Heffner and M.R. Norman, Comm. Condens. Matter Phys.
{\bf 17}, 361 (1996).
\bibitem{Aeppli89} G. Aeppli {\it et al.\/}, Phys. Rev. Lett. {\bf 60}, 615
(1989).
\bibitem{Isaacs95} E.D. Isaacs {\it et al.\/}, Phys. Rev. Lett. {\bf 75}, 1178
(1995).
\bibitem{Tou} H. Tou {\it et al.\/}, Phys. Rev. Lett. {\bf 77}, 1374 (1996).
\bibitem{Dalmas95} P. Dalmas de R\'eotier {\it et al.\/}, Phys. Lett. A 
{\bf 205}, 239 (1995).
\bibitem{extra} The increase of the muon damping rate observed below $T_N$ 
with a first generation sample 
(e.g. D.W. Cooke {\it et al.\/}, Hyperfine Interact. {\bf 31}, 425 (1986)
and R.H. Heffner {\it et al.\/}, Phys. Rev. B {\bf 39}, 11345 (1989))
has not been reproduced thereafter (see
Refs. \protect\onlinecite{Dalmas95,Visser97b}).
\bibitem{muon} A. Schenck, and F.N. Gygax, in {\sl Handbook of Magnetic 
Materials\/}, Vol. 9, edited by K.H.J. Buschow (Elsevier Science B.V., 1995).
; P. Dalmas de R\'eotier and A. Yaouanc, J. Phys.: Condens.
Matter {\bf 9}, 9113 (1997); A. Amato, Rev. Mod. Phys. {\bf 69}, 1119 (1997).
\bibitem{Yaouanc98} A. Yaouanc  {\it et al.\/}, J. Phys.: Condens. Matter 
{\bf 10}, 9791 (1998).
\bibitem{Suderow98} H. Suderow {\it et al.\/}, Phys. Rev. Lett. {\bf 80}, 165
(1998).
\bibitem{Abela99} R. Abela {\it et al.\/}, Hyperfine Interact. {\bf 120-121},
575 (1999).
\bibitem{Schenck82} A. Schenck, Helv. Phys. Acta {\bf 54}, 471 (1982).
\bibitem{Lussier96} B. Lussier {\it et al.\/}, Phys. Rev. B {\bf 54}, R6873
(1996).
\bibitem{Dijk98} N.H. van Dijk {\it et al.\/}, Phys. Rev. B {\bf 58}, 
3186 (1998).  
\bibitem{Frings83} P.H. Frings {\it et al.\/}, J. Magn. Magn. Mat. {\bf 31-34},
240 (1983).
\bibitem{Schenckfuture} A. Schenck {\it et al.\/}, cond-mat/9909197.   
\bibitem{Visser97b} A. de Visser {\it et al.\/}, Physica B {\bf 230-232}, 53
(1997).
\bibitem{Midgley93} P.A. Midgley {\it et al.\/}, Phys. Rev. Lett. {\bf 70},
678 (1993).

 
\end{references}
\end{document}